# Hall and field-effect mobilities in few layered $p$-WSe$_2$ field-effect transistors

N. R. Pradhan[1], D. Rhodes[1], S. Memaran[1], J. M. Poumirol[1], D. Smirnov[1], S. Talapatra[2], S. Feng[3], N. Perea-Lopez[3], A. L. Elias[3], M. Terrones[3], P. M. Ajayan[4] & L. Balicas[1]

[1]*National High Magnetic Field Laboratory, Florida State University, Tallahassee-FL 32310, USA*

[2]*Physics Department, Sourthern Illinois University, Carbondale-IL 62901-4401, USA.*

[3]*Department of Physics, Department of Materials Science and Engineering and Materials Research Institute, The Pennsylvania State University, University Park, PA 16802, USA.*

[4]*Department of Mechanical Engineering and Materials Science, Rice University, Houston, TX 77005 USA*

balicas@magnet.fsu.edu

Here, we present a temperature ($T$) dependent comparison between field-effect and Hall mobilities in field-effect transistors based on few-layered WSe$_2$ exfoliated onto SiO$_2$. *Without* dielectric engineering and beyond a $T$-dependent threshold gate-voltage, we observe maximum hole mobilities approaching 350 cm$^2$/Vs at $T$=300 K. The hole Hall mobility reaches a maximum value of 650 cm$^2$/Vs as $T$ is lowered below ~150 K, indicating that insofar WSe$_2$-based field-effect transistors (FETs) display the largest Hall mobilities among the transition metal dichalcogenides. The gate capacitance, as extracted from the Hall-effect, reveals the presence of spurious charges in the channel, while the two-terminal sheet resistivity displays two-dimensional variable-range hopping behavior, indicating carrier localization induced by disorder at the interface between WSe$_2$ and SiO$_2$. We argue that improvements in the fabrication protocols as, for example, the use of a substrate free of dangling bonds are likely to produce WSe$_2$-based FETs displaying higher room temperature mobilities, i.e. approaching those of *p*-doped Si, which would make it a suitable candidate for high performance opto-electronics.

Field-effect transistors (FETs) based on exfoliated transition-metal dichalcogenides (TMDs)[1-4] were shown to be promising as low-power switching devices and therefore as potential components for high-resolution liquid crystal and organic light-emitting diode displays, particularly in their multi-layered form.[5] Bulk transition metal dichalcogenides (TMD) crystallize in the "2$H$" or trigonal prismatic structure (space group $P6_3/mmc$), in which each transition metal is surrounded by six chalcogenide atoms defining two triangular prims. Extended planes, which are weakly or van der Waals coupled, result from the tessellation of this basic unit. Contiguous planes are shifted with respect to one another (along both the **a-** and the **b-**axis), therefore the unit cell is composed of two planes with a transition metal dependent inter-layer distance $c$. The covalently bonded layers are expected to display high crystallinity, although as in graphite/graphene, one can expect crystallographic mosaicity between planes stacked along the $c$-axis. Similarly to graphite, compounds such as $MoS_2$, $WS_2$, etc., are exfoliable layered materials characterized by a weak inter-planar van der Waals coupling.[1] In contrast to graphene, they exhibit indirect band gaps ranging from ~ 1 to ~ 2 eV which become direct in single atomic-layers,[6] making them promising candidates for applications.

Early studies[7] on field-effect transistors (FETs) based on bulk $WSe_2$ single-crystals using parylene as the gate dielectric, revealed room temperature field-effect mobilities approaching those of $p$-Si[8] but with a small current ON/OFF ratio. Subsequent investigations[9] on mechanically exfoliated $MoS_2$ flakes composed of tenths of atomic layers and $SiO_2$ as the gate dielectrics, revealed considerably lower mobilities (10 - 50 cm$^2$/Vs), suggesting either a remarkable difference in mobilities between $MoS_2$ and $WSe_2$ or that an inadequate choice of gate dielectrics can hinder their performance. More recently,[10] it was suggested that field-effect carrier mobilities surpassing 1000 cm$^2$/Vs could be achieved in dual gated, single-layer $MoS_2$

FETs through the use of a top gate composed of a high-κ dielectric such as HfO$_2$. Nevertheless, it was argued that this is an overestimated mobility value due to the capacitive coupling between both top and back gates,[11] a fact that is supported by subsequent reports of much smaller mobilities in similar devices when the gate capacitance is extracted from a Hall-effect study.[12,13] It was also recently argued that remote phonons from dielectric layers such as HfO$_2$, can limit carrier mobility and would require the use of an interfacial layer to absorb most of the vibrational energy.[14] Nevertheless, these observations already led to the development of integrated circuits based on single[15]- and on bi-layered[16] MoS$_2$. Recent studies in both single- and double- layered MoS$_2$ revealed Hall mobilities which increase strongly with gate voltage, saturating at maximum values between ~ 200 and ~ 375 cm$^2$/Vs at low temperatures.[17] In multi-layered MoS$_2$ the Hall mobility has been found to increase from ~ 175 cm$^2$/Vs at 60 K to 311 cm$^2$/Vs at $T = 1$ K at back-gate voltages as large as 100 V.[18] However, marked discrepancies were reported between the measured field-effect and the Hall mobilities,[17] which at the light of Refs.[11-13] could be attributed to underestimated values for the gate capacitances.

Similarly to past research on graphene, much of the current effort on TMD-based FETs is focused on understanding the role played by the substrates, annealing conditions and the work functions of the metallic contacts. For example, it was recently argued that most of the above quoted mobilities are determined by the Schottky barriers at the level of the current contacts which limits the current-density that can be extracted from these transistors. The authors of Ref.[19] argue that small Schottky barriers, and therefore nearly Ohmic contacts in TMD based FETs, can only be achieved through the use of metals with small work functions such as Sc. Furthermore, due to the detrimental role played by the SiO$_2$ substrates, Ref.[19] finds that the highest mobilities (~ 175 cm$^2$/Vs) can be achieved in FETs built on ~ 10 nm (~ 15 layers) thick flakes. Thickness dependent mobilities were also recently reported for $MoS_2$ based transistors using polymethyl methacrylate (PMMA) as the gate dielectrics.[20] High performance TMD-based FETs have been claimed to have the potential to make a major impact in low power optoelectronics.[5,21-23] Here, to evaluate this assertion, we study and compare field-effect and Hall mobilities in field-effect transistors based on few-layered $WSe_2$ exfoliated onto $SiO_2$, finding that they can display room temperature hole-mobilities approaching those of hole-doped Si[8] with a large ON to OFF ratio ($> 10^6$) and sharp subthreshold swings (ranging from 250 and 140 mV per decade). This observation is remarkable given that i) carrier mobility is expected to be limited by the scattering from intrinsic[24] as well as substrate phonons, ii) the Schottky barriers at the contacts have yet to be optimized, and as we show iii) the presence of charge traps and disorder at the interface between $WSe_2$ and $SiO_2$ should limit the carrier mobility. Improvements in device fabrication, can lead to improved performance with respect to these values open promising prospects for optoelectronic applications.

**Results and Discussion**

Figures 1 **a** and **b** show respectively, a micrograph of a typical device, whose experimental results will be discussed throughout this manuscript, and the sketch of a four-terminal configuration for conductance measurements. Current source $I^+$ and drain $I^-$ terminals, as well as the pairs of voltage contacts 1, 2 and 3, 4 are indicated. As shown below, this configuration of contacts allows us to compare electrical transport measurements performed when using a 2-contact configuration (e.g. $\mu_{FE}$) with a 4-terminal one (e.g. $R_{xy}$ or the Hall-effect). Figure 1 **c** shows an atomic force microscopy profile and image (inset) from which we extract a flake thickness of ~ 8 nm, or approximately 12 atomic layers. We chose to focus on multi-layered FETs because our *preliminary* observations agree with those of Refs.[19,20], indicating that the highest mobilities are observed in flakes with thicknesses between ~ 10 and 15 atomic layers as shown in Fig. 1 **d**. In addition, as argued in Ref.[5] multilayered flakes should lead to thin film transistors yielding higher drive currents when compared to transistors based on single atomic layers, possibly making multilayered FETs more suitable for high-resolution liquid crystal and organic light-emitting diode displays.[5] Our flakes were mechanically exfoliated and transferred onto a 270 nm thick $SiO_2$ layer grown on *p*-doped Si, which is used as a back gate. Throughout this study, we focus on devices with thicknesses ranging from 9 to 15 layers. Three of the devices were annealed at 150 K under high vacuum for 24 h, which as reported in Ref.[17], yields higher mobilities particularly at low temperatures. We found very similar overall response among the non-annealed samples, as well as among the annealed ones.

Figure 2 **a** shows the extracted field-effect current $I_{ds}$ as a function of the back gate voltage $V_{bg}$ for several fixed values of the voltage $V_{ds}$ across the current contacts, i.e. when using a 2-terminal configuration. From initial studies,[7] but in contrast with Refs.[25,26], $WSe_2$ is expected to show ambipolar behavior, i.e. a sizable current resulting from the accumulation of either electrons or holes at the $WSe_2/SiO_2$ interface due to the electric field-effect. Although we have previously observed such a behavior, all FETs studied here show a rather modest electron current (i.e. saturating at ~$10^{-8}$ A) at positive $V_{bg}$ values in contrast also with samples covered with $Al_2O_3$, see Ref.[26]. Therefore our samples behave as if hole-doped (i.e. sizeable currents only for negative gate voltages). At room temperature the minimum current is observed around $V_{bg} \approx 0$ V while the difference in current between the transistor in its "ON"-state with respect to the OFF-one (on/off ratio) is $> 10^6$. For all measurements, the maximum channel current was limited in order to prevent damaging our FETs. The subthreshold swing SS is found to be ~250 mV per decade, or ~3.5 times larger than the smallest values extracted from Si MOSFETs at room

temperature. Figure 2**b** shows the conductivity $\sigma = I_{ds}\ell/V_{ds}w$ (from **a**), as a function of $V_{bg}$ for several values of $V_{ds}$. As indicated in the caption of Fig. 1 the separation between the current contacts, is $\ell = 15.8$ μm while the width of the channel is $w = 7.7$ μm. As seen, all curves collapse on a single curve indicating linear behavior, despite the claimed role for Schottky barriers at the level of contacts.[19] See also the Supplemental Information section for linear current-voltage characteristics for the range of excitation voltages used. Figure 2 **c**: the field-effect mobility $\mu_{FE}$ can be evaluated in the standard way by normalizing by the value of the gate capacitance ($c_g = 12.789 \times 10^{-9}$ F/cm$^2$) the derivative of the conductivity with respect to $V_{bg}$. As seen, $\mu_{FE}$ increases sharply above $V_{bg} \approx 2$ V reaching a maximum of ~305 cm$^2$/Vs at $V_{bg} \sim -20$ V, decreasing again beyond this value. Alternatively, the mobility can be directly evaluated through the slope of $I_{ds}$ as a function of $V_{bg}$ in its linear regime, and by normalizing it by the sample geometrical factors, the excitation voltage $V_{bg}$ and the gate capacitance $c_g$, yielding a peak value $\mu_{FE} \approx 302$ cm$^2$/Vs. We have observed $\mu_{FE}$ values as high as 350 cm$^2$/Vs (see results for sample 2 below). These values, resulting from two-terminal measurements, are comparable to those previously reported by us in multi-layered MoS$_2$, where we used a four-terminal configuration to eliminate the detrimental role played by the less than ideal contacts.[27]

Figures 3 **a**, **b**, **c**, and **d** show respectively, $I_{ds}$ as a function of $V_{bg}$ for several values of $V_{ds}$, the corresponding conductivities σ as a function of $V_{bg}$, and the resulting field-effect mobility as previously extracted through Figs. 2 **c** and **d**. All curves were acquired at $T = 105$ K. As seen, at lower temperatures $\sigma(T, V_{bg})$ still shows a linear dependence on $V_{ds}$ although lower $T$s should be less favorable for thermally activated transport across Schottky barriers. In fact, we collected similarly linear data sets at $T < 105$ K. At $T = 105$ K, $\mu_{FE}$ displays considerably higher values, i.e. it surpasses 650 cm$^2$/Vs (accompanied by reduction in the SS down to ~140 mV per

decade). However, as seen in Fig. 3 **a**, lower temperatures increase the threshold gate voltage $V^t_{bg}$ required for carrier conduction. Below we argue that this is the result of a prominent role played by disorder and/or charge traps at the interface between $WSe_2$ and $SiO_2$ instead of just an effect associated with the Schottky barriers. Large Schottky barriers are expected to lead to non-linear current $I_{ds}$ as a function of the excitation voltage $V_{ds}$ characteristics, with a sizeable $I_{ds}$ emerging only when $V_{ds}$ surpasses a threshold value determined by the characteristic Schottky energy barrier ϕ, as seen for instance in Ref.[28]. But according to Figs. 2 **b** and 3 **b**, σ is basically independent on $V_{ds}$ above a threshold gate voltage, *even at lower temperatures*.

Figure 4 **a** shows $I_{ds}$ as a function of $V_{bg}$ for several temperatures and for the crystal shown in Fig. 1 **a**. Fig. 4 **b** shows the resulting field-effect mobility $\mu_{FE}$ as a function of $T$ as extracted from the slopes of $I_{ds}(V_{bg}, T)$. $\mu_{FE}$ is observed to increase, reaching a maximum of ~ 650 cm$^2$/Vs at $T$ ~100 K, decreasing subsequently to values around 250 cm$^2$/Vs at low temperatures. Orange markers depict $\mu_{FE}$ for a second, annealed sample whose Hall mobility is discussed below. This decrease is attributable to extrinsic factors, such as chemical residues from the lithographic process, since annealing the samples under high vacuum for at least 24 h considerably increases the mobility at low $T$s,[17] as will be illustrated by the results shown below for a second sample annealed in this way. Figure 4 **c** shows $\mu_{FE}$ as a function of $V_{bg}$ for several temperatures (as extracted from the curves in **a**). All curves show a maximum at a $V_{bg}$-dependent value. As seen, the main effect of lowering $T$ is to increase the threshold back-gate voltage $V^t_{bg}$ for carrier conduction. In $WS_2$, by using ambipolar ionic liquid gating, which heavily screens charged defects, the authors of Ref.[29] were able to estimate the size of its semiconducting gap, given roughly by the difference between the threshold voltages required for hole and electron conduction respectively, or ~1.4 V. The much larger $V^t_{bg}$ values observed by us in $WSe_2$ is attributable to intrinsic and extrinsic effects, such as vacancies and charge traps, which limit the carrier mobility becoming particularly relevant at low temperatures, see discussion below. At first glance, at low gate voltages ρ would seem to follow activated behavior with a small activation gap. On the other hand at high temperatures and high gate voltages, ρ displays metallic like behavior, usually defined by $\partial \rho/\partial T > 0$. Magenta line is a fit to a simple linear-dependence on temperature, suggesting either an unconventional metallic state or most likely, phonon scattering.

As observed in Figs. 4 **a** and **c**, the threshold gate-voltage $V^t_{bg}$ required to observe a finite σ increases from ~ 5 to ~ 35 V as $T$ is lowered from 300 to 5 K. In order to clarify the dependence of $V^t_{bg}$ on $T$, we assume that $V^t_{bg}$ is dominated by disorder at the interface between $WSe_2$ and $SiO_2$ which leads to charge localization. To illustrate this point, in Fig. 5 we plot σ($T$) as function of $T^{-1/3}$ since from past experience on Si/SiO$_2$ MOSFETs, it is well known that spurious charges intrinsic to the SiO$_2$ layer,[30-32] in addition to the roughness at the interface between the Si and the glassy SiO$_2$,[33] produces charge localization leading to variable-range hopping conductivity: $\sigma(T) = \sigma_0 \exp(-T_0/T)^{1/(1+d)}$ where $d$ is the dimensionality of the system, or $d$ = 2 in our case.[34] As seen in Fig. 5, one observes a crossover from metallic-like to a clear two-dimensional variable-range hopping (2DVRH) conductivity below a gate voltage dependent temperature; red lines are linear fits. At lower gate voltages, the 2DVRH regime is observed over the entire range of temperatures. Therefore, despite the linear transport regime and the relatively large mobilities observed in Figs. 1 through 4, this plot indicates very clearly, that below $V^t_{bg}$ the carriers in the channel are localized due to disorder. Notice that similar conclusions were also reported from measurements on MoS$_2$.[35] Although, at the moment we do not have a clear experimental understanding on type and on the concomitant role of disorder in these systems

(which would allow a deeper theoretical understanding on the origin of the localization), the above experimental plot is unambiguous in revealing the predominant conduction mechanism for gate-voltages below a threshold value.

Now, we are in position of qualitatively explaining the $T$-dependence of $V^t_{bg}$: thermal activated processes promote carriers across a mobility edge which defines the boundary between extended electronic states and a tail in the density of states composed of localized electronic states. At higher temperatures, more carriers are thermally excited across the mobility edge, or equivalently, can be excited across the potential well(s) produced by disorder or charge traps, therefore one needs lower gate voltage(s) to untrap the carriers. Once these carriers have moved across the mobility edge, they become mobile and, as our results show, respond linearly as a function of the excitation voltage $V_{ds}$. Finally, as $V^t_{bg}$ increases with decreasing $T$ the number of carriers is expected to *decrease* continuously since they become progressively localized due to the suppression of thermally activated processes which can no longer contribute to carrier detrapping. This is clearly illustrated by Fig. 4 **b**, where one sees an increase in mobility, due to the suppression of phonon scattering, leading to a maximum in the mobility and to its subsequent suppression upon additional cooling. Therefore, at higher temperatures and for gate voltages above the threshold, where one observes a metallic-like state, one has two competing mechanisms at play upon cooling, i.e. the tendency to localization/suppression of carriers which is unfavorable to metallicity, and the suppression of phonon scattering. Suppression of phonon scattering is the only possible explanation for the observed metallic behavior. Hence, one must conclude that this metallic behavior ought to be intrinsic to the compound, but disorder-induced carrier localization dominates σ at lower temperatures.

Although, as Figs. 2 and 3 indicate, the conductivity σ as measured through a two-terminal configuration, is linear on excitation voltage $V_{ds}$ when $V_{bg} > V^t_{bg}$, it was discussed at length that the electrical conduction through the drain and source contacts can by no means be ohmic.[19,36] In effect, a Schottky barrier of ~770 meV is expected as the difference in energy between the work function of Ti, or 4.33 eV, and the ionization energy of WSe$_2$, or ~ 5.1 eV.[37,38] The linear, or apparent ohmic regime presumably would result from thermionic emission or thermionic field emission processes. According to thermionic emission theory, the drain-source current $I_{ds}$ is related to the Schottky barrier height $\phi_{SB}$ through the expression:

$$I_{ds} = AA^*T^2 \exp(e\phi_{SB}/k_BT) \quad (1)$$

Where $A$ is the area of the Schottky junction, $A^* = 4\pi em^*k_B^2h^{-3}$ is the effective Richardson constant, $e$ is the elementary charge, $k_B$ is the Boltzmann constant, $m^*$ is the effective mass and $h$ is the Planck constant.[39] In order to evaluate the Schottky barrier at the level of the contacts, in the top panel of Fig. 6 we plot $I_{ds}$ normalized by the square of the temperature $T^2$ as a function of $e/k_BT$ and for several values of the gate voltage. Red lines are linear fits from which we extract the $\phi_{SB}(V_{bg})$. Notice that in the top panel of Fig. 6 the linear fits are limited to higher temperatures since at lower temperatures one observes pronounced, gate dependent, deviations from the thermionic emission theory. The bottom panel of Fig. 6 shows $\phi_{SB}(V_{bg})$ in a logarithmic scale as a function of $V_{bg}$. Red line is a linear fit from whose deviation we extract the size of the Schottky barrier,[19] or Φ ~ 16 meV, indicating a much better band alignment than originally expected. It is perhaps possible that the Eq. (1) might take a different form for layered two-dimensional materials, for example, in such compounds one might need a temperature pre-factor distinct from $T^2$. We attempted the use of different temperature pre-factors such as $T$ or $T^{3/2}$, but it does not improve the linearity of $\log(I_{ds}/T^\alpha)$ (with $2 \geq \alpha \geq 1$ ) as a function of $ek_B/T$. In fact, an arbitrary $T$ pre-factor, would not be theoretically justifiable at the moment. Having said that, one has to be very careful with the extraction of the Schottky barrier through this common approach, since the two-terminal measurements contain contributions from both the contacts and the conduction channel which, as discussed above, underdoes disorder-induced carrier localization, thus masking the true behavior of the conduction across the contacts. Notice for example, how in Fig. 5 the 2DVRH fits the behavior of the $\sigma(T)$ over the entire range of temperatures when $V_{bg}$ = - 20 V, while in Fig. 6, it can describe the behavior of $I_{ds}/T^2$ as a function of $T^{-1}$ only when $T >$ 125 K. Therefore the values of $\phi_{SB}(V_{bg})$ extracted here should be taken with caution.

In Figure 7, we compare the above field-effect mobilities with Hall mobility measurements on a second, vacuum annealed flake of similar thickness. Figure 7 **a** shows the four-terminal sheet resistivity, i.e. $\rho_{xx} = wV_{ds}/lI_{ds}$ as a function of $V_{bg}$. $\rho_{xx}$ was measured with a lock-in technique, for gate voltages where the voltages $V_{12}$ or $V_{34}$ were in phase with the excitation signal. We also checked that any pair of voltage contacts produced nearly the same value for $\rho_{xx}$, indicating a nearly uniform current throughout the channel. $\rho_{xx}$ increases very rapidly, beyond $10^9$ Ω as $V_{bg} \rightarrow 0$ V. Also the out-of-phase component of the measured AC signal becomes very large as $V_{bg} \rightarrow 0$ limiting the $V_{bg}$ range for our measurements. Figure 7 **b** displays the measured Hall signal $R_{xy}$ as a function of the magnetic field $H$ at $T = 50$ K and for several values of $V_{bg}$. Red lines are linear fits from which we extract the Hall constant $R_H = R_{xy}/H$ = $1/ne$. In the same Fig. 7 **b** we also indicate the extracted values for the Hall mobilities, $\mu_H =$ $R_H/\rho_{xx}$, at different gate voltages. Notice that for $T = 50$ K and $V_{bg} = 70$ V one obtains, in this annealed sample, a $\mu_H$ value of ~676 cm$^2$/Vs. Figure 7 **c** shows the density of carriers $n_H = 1/eR_H$ as a function of $V_{bg}$ for several $T$s. Red lines are linear fits from which we extract the slope $n_H/V_{bg} = c_g^*/e$, where $c_g^*$ is an effective back-gate capacitance: in the absence of extrinsic charged defects at the WSe$_2$/SiO$_2$ interface, $c_g^*$ should be equal to the previously quoted gate capacitance $c_g$. Solid evidence for the existence of ionized impurities acting as hole traps at the interface is provided by the linear fits in Fig. 7 **c** which intercepts the $n_H = 0$ axis at finite threshold gate voltages $V_{bg}^t$. This confirms that practically all holes generated by applying a gate voltage smaller than $V_{bg}^t$ remain localized at the interface. Figure 7 **d** shows a comparison between $\mu_{FE}$ (magenta and blue lines) and $\mu_H$ (red markers) as extracted from the same device at room temperature. The blue line was measured after thermally cycling the FET down to low temperatures. Notice how $V_{bg}^t$ increases after thermally cycling the sample, thus suggesting that strain at the interface, resulting from the difference between the thermal expansion coefficients of SiO$_2$ and WSe$_2$, also contributes to $V_{bg}^t$. Therefore, strain would seem to be an additional factor contributing to the mobility edge. Notice also that both mobilities initially increase as a function $|V_{bg}|$, reaching a maximum at the same $V_{bg}$ value, decreasing subsequently as the back-gate voltage is further increased. Figure 7 **e** shows $\mu_H$ as a function of $T$ for several values of $V_{bg}$. Notice how $\mu_H$ ($T \rightarrow 0$ K) is suppressed at low gate voltages due to the charge localization mechanism discussed above. $\mu_H$ is observed to increase as $T$ is lowered, requiring ever increasing values of $V_{bg} > V_{bg}^t$, but decreases again below $T \sim 5$ K. A fit of $\mu_H(T, V_{bg} = -60$ V$)$ to $AT^{-\alpha}$ yields $\alpha \sim (1 \pm 0.1)$. Finally Fig. 7 **f** displays the $T$-dependence of the ratio between the measured and the ideal geometrical gate capacitance ($c_g^* = se$)/$c_g$ where $s$ corresponds to the slopes extracted from the linear-fits in Fig. 7 **c**. For a perfect FET this ratio should be equal to 1, i.e. the only charges in the conducting channel should be those resulting from the electric field-effect. Therefore, one can estimate the carrier mobility $\mu_i$ for the nearly ideal device, i.e. with the ideal geometrical capacitance, through $\mu_i = c_g^*/c_g\,\mu_H$, which at $T = 300$ K would lead to $V_{bg}$-dependent mobilities ranging from 350 up to 525 cm$^2$/Vs. This rough estimate does not take into account scattering processes resulting from for example, other sources of disorder within the channel. In agreement with Ref.[40], this indicates that in our $WSe_2$ FETs the main scattering mechanism limiting the carrier mobility are not phonons, but ionized impurities and disorder, or that phonon scattering would still allow mobilities approaching, and probably surpassing, 500 cm$^2$/Vs at room temperature. In *p*-doped Si the hole-mobility is observed to saturate at a value of ~475 cm$^2$/Vs for doping levels below ~$10^{17}$ per cm$^3$, while a doping concentration of $10^{19}$ per cm$^3$ yields mobilities of ~200 cm$^2$/Vs as observed here.[8] Therefore, our work indicates that if one was able to improve the FET fabrication protocols, by minimizing the disorder such as interface roughness, spurious ionized impurities and dangling bonds at the interface, $WSe_2$ could match the performance of *p*-doped Si, thus becoming suitable for specific applications[5] with the added advantage of miniaturization, since the starting point would be just a few atomic layers.

Notice that the $\mu_{FE}$ values extracted here at higher *T*s would be overestimated if one considers the value of the gate capacitance extracted from the Hall effect, i.e. it would be two to three times larger than the expected geometrical capacitance, thus implying 2 to 3 times smaller values for $\mu_{FE}$. A number of reports on TMDs[16,19,20] suggest room temperature field-effect mobilities ranging from 300 to ~700 cm$^2$/Vs for $MoS_2$ based FETs subjected to "dielectric engineering". However, taken together with the debate in Refs.[11,12] concerning the true value of the gate capacitance in dual gated FETs, our study suggests that those values should be carefully re-examined by performing four-terminal Hall-mobility and/or capacitance measurements.

In the Supplemental Information, we show the Raman spectra of $WSe_2$ whose main Raman modes are observed to sharpen considerably as the number of layers decrease, implying a pronounced increase in phonon lifetimes. Possibly, the main source of disorder in $WSe_2$ is stacking disorder, which is progressively eliminated as one decreases the number of layers. This also implies a high degree of in-plane crystallinity. On the other hand, polarized Raman indicates that most Raman modes in WSe$_2$ are mixed modes, i.e. composed of in-plane and out-of-plane lattice vibrations, which might affect the strength of its electron-phonon coupling.

Although a gate-voltage dependent Raman study has yet to be performed in WSe$_2$, in both single-layer[41] and bi-layer[42] graphene, it was observed that the gate-voltage can tune the interaction between phonons and the charge carriers, leading to changes in the amplitude and in the line-width of the Raman spectra. A similar gate-voltage dependence in WSe$_2$ might reveal reduced electron-phonon scattering therefore explaining the higher room-temperature Hall mobilities observed here. Notice, that monolayer TMDs have been predicted to display strong piezoelectricity,[43] suggesting that these materials are prone to a strong coupling between lattice degrees of freedom and an external electric field.

**Conclusions**

In summary field-effect transistors based on multi-layered *p*-doped WSe$_2$ can display peak hole Hall-mobilities in excess of 200 cm$^2$/Vs at room temperature. This value increases by a factor > 3.3 when the temperature decreases to ~ 100 K. The carrier density as a function of the gate voltage, as extracted from the Hall-effect, indicates larger than expected gate capacitances thus implying an excess of spurious charges in the channel. Therefore, one should be cautious when quoting values for the field-effect mobility by using the geometrical gate capacitance value. These spurious charges, in addition to disorder at the WSe$_2$/SiO$_2$ interface, leads to carrier localization and to a concomitant mobility edge, which manifests itself in an increasing threshold gate voltage for carrier conduction and, at a fixed gate voltage, to a concomitant decrease in carrier mobility upon cooling (resulting from an increase in the threshold gate voltage). When using Ti:Au for the electrical contacts one obtains a remarkable small value for the size of the

Schottky barrier, although thermionic emission theory can only properly fit the transport data at higher temperatures.

We emphasize that our results indicate that WSe$_2$ displays what seemingly are the highest Hall mobilities observed so far in TMDs, particularly among FETs based on few-layered TMDs exfoliated onto SiO$_2$ and remarkably, without the use of distinct or additional dielectric layers. The Hall mobility values observed here surpass, for example, the μ$_H$ values in Ref.[17] for MoS$_2$ on HfO$_2$ or the field-effect mobilities of thicker multilayered MoS$_2$ flakes[5] on Al$_2$O$_3$. This indicates that WSe$_2$ has the potential to display even higher carrier mobilities, particularly at room temperature, through the identification of suitable substrates (flatter interfaces, absence of impurities and dangling bonds, etc), as well as contact materials. A major materials research effort must be undertaken to clarify the density of point defects (e.g. vacancies, intercalants) in the currently available material and on how to decrease their density. However, our study reveals that WSe$_2$ has the potential to become as good if not a better material for optoelectronic applications than, for instance, multi-layered MoS$_2$.[5] Recently, Ref.[44] reported the performance of multi-layered WSe$_2$ FETs, composed of WSe$_2$ atomic layers transferred onto a *h*-BN substrate using graphene for the electrical contacts as well as ionic liquid gating. Remarkably, despite the complexity of this architecture, originally intended to improve the overall performance of multi-layered WSe$_2$ FETs, the simpler devices reported here, still display considerably higher mobilities. We believe this is an important piece of information for those considering the development of electronic or optoelectronic applications based on transition metal dichalcogenides.

**Methods**

WSe$_2$ single crystals were synthesized through a chemical vapor transport technique using iodine as the transport agent. Multi-layered flakes of WSe$_2$ were exfoliated from these single crystals by using the "scotch-tape" micromechanical cleavage technique, and transferred onto *p*-doped Si wafers covered with a 270 nm thick layer of SiO$_2$. Prior to transferring the WSe$_2$ crystals onto the SiO$_2$ layers, these were cleaned in the following way: SiO$_2$ was sonicated for 15 min in acetone, isopropanol and deionized water, respectively. It was subsequently dried by a nitrogen gas flow. For making the electrical contacts 90 nm of Au was deposited onto a 4 nm layer of Ti *via* e-beam evaporation. Contacts were patterned using standard e-beam lithography techniques. After gold deposition, the devices were annealed at 200 °C for ~2 h in forming gas. Atomic force microscopy (AFM) imaging was performed using the Asylum Research MFP-3D AFM. Electrical characterization was performed by using a combination of sourcemeter (Keithley 2612 A), Lock-In amplifier (Signal Recovery 7265) and resistance bridges (Lakeshore 370) coupled to a Physical Property Measurement System. The Raman spectra were measured in a backscattering geometry using a 532.1 nm laser excitation. For additional details see the Supplemental Information.

1. Wang, Q. H., Kalantar-Zadeh, K., Kis, A., Coleman, J. N., & Strano, M. S. Electronics and Optoelectronics of Two-Dimensional Transition Metal Dichalcogenides. *Nat. Nanotechnol.* **7**, 699 (2014).

2. Chhowalla, M. *et al*. The Chemistry of Two-Dimensional Layered Transition Metal Dichalcogenide Nanosheets. *Nat. Chem.* **5**, 263-275 (2013).

3. Zeng, H. L., Dai, J. F., Yao, W., Xiao, D., & Cui, X. D. Valley Polarization in MoS$_2$ Monolayers by Optical Pumping. *Nat. Nanotechnol.* **7**, 490-493 (2012).

4. Mak, K. F., He, K. L., Shan, J., & Heinz, T. F. Control of Valley Polarization in Monolayer MoS$_2$ by Optical Helicity. *Nat. Nanotechnol.* **7**, 494-498 (2012).

5. Kim S. *et al*. High-Mobility and Low-Power Thin-Film Transistors Based on Multilayer MoS$_2$ Crystals. *Nat. Commun.* **3**, 1011 (2012).

6. Tonndorf, P. *et al*. Photoluminescence Emission and Raman Response of Monolayer MoS$_2$, MoSe$_2$, and WSe$_2$. *Opt. Express* 21, 4908-4916 (2013)

7. Podzorov, V., Gershenson, M. E., Kloc, Ch., Zeis, R., & Bucher, E. High-Mobility Field-Effect Transistors Based on Transition Metal Dichalcogenides. *Appl. Phys. Lett.* **84**, 3301 (2004).

8. Reggiani, S. *et al*. Electron and Hole Mobility in Silicon at Large Operating Temperatures - Part I: Bulk mobility. *IEEE T. Electron. Dev.* **49**, 490 (2002).

9. Ayari, A., Cobas, E., Ogundadegbe, O., & Fuhrer, M. S. Realization and Electrical Characterization of Ultrathin Crystals of Layered Transition-Metal Dichalcogenides. *J. Appl. Phys.* **101**, 014507 (2007).

10. Lembke, D., & Kis, A. Breakdown of High-Performance Monolayer MoS$_2$ Transistors. *ACS Nano* **6**, 10070-10075 (2012).

11. Fuhrer, M. S., & Hone, J. Measurement of Mobility in Dual-Gated MoS$_2$ Transistors. *Nat. Nanotechnol.* **8**, 146-147 (2012).

12. Radisavljevic, B., & Kis, A. Measurement of Mobility in Dual-Gated MoS$_2$ Transistors. *Nat. Nanotechnol.* **8**, 147-148 (2013).

13. Radisavljevic, B., & Kis, A. Mobility Engineering and a Metal-Insulator Transition in Monolayer MoS$_2$. *Nat. Mater.* **12**, 815 (2013).

14. Zeng, L. *et al*. Remote Phonon and Impurity Screening Effect of Substrate and Gate Dielectric on Electron Dynamics in Single Layer MoS$_2$. *Appl. Phys. Lett.* **103**, 113505 (2013)

15. Radisavljevic, B., Whitwick, M. B., & Kis, A. Integrated Circuits and Logic Operations Based on Single-Layer MoS$_2$. *ACS Nano* **12**, 9934 (2011).

16. Wang, H. *et al*. Integrated Circuits Based on Bilayer MoS$_2$ Transistors. *Nano Lett.* **12**, 4674 (2012).

17. Baugher, B. W. H., Churchill, H. O. H., Yang, Y., Jarillo-Herrero, P. Intrinsic Electronic Transport Properties of High-Quality Monolayer and Bilayer MoS$_2$. *Nano Lett.* **13**, 4212-4216 (2013).

18. Neal, A. T., Liu, H., Gu, J., & Ye, P. D. Magneto-transport in MoS$_2$: Phase Coherence, Spin-Orbit Scattering, and the Hall Factor. *ACS Nano* **8**, 7077-7082 (2013).

19. Das, S., Chen, H.-Y., Penumatcha, A. V., & Appenzeller, J. High Performance Multilayer MoS$_2$ Transistors with Scandium Contacts. *Nano Lett.* **13**, 100-105 (2013).

20. Bao, W., Cai, X., Kim, D., Sridhara, K., & Fuhrer, M. S. High Mobility Ambipolar MoS$_2$ Field-effect Transistors: Substrate and Dielectric Effects. *Appl. Phys. Lett.* **102**, 042104 (2013).

21. Yin, Z. *et al*. Single-Layer MoS$_2$ Phototransistors. *ACS Nano* **6**, 74 (2012).

22. Lee, H. S. *et al*. MoS$_2$ Nanosheet Phototransistors with Thickness-Modulated Optical Energy Gap. *Nano Lett.* **12**, 3695 (2012).

23. Choi, W. *et al*. High-Detectivity Multilayer MoS$_2$ Phototransistors with Spectral Response from Ultraviolet to Infrared. *Adv. Mater.* **43**, 5832 (2012).

24. Kaasbjerg, K., Thygesen, K. S., & Jacobsen, K. W. *Phys. Rev. B* **85**, 115317 (2012).

25. Fang, H. *et al*. High-Performance Single Layered WSe$_2$ *p*-FETs With Chemically Doped Contacts. *Nano Lett.* **12**, 3788 (2012).

26. Liu, W. *et al*. Role of Metal Contacts in Designing High-Performance Monolayer *n*-Type WSe$_2$ Field Effect Transistors. *Nano Lett.* **13**, 1983-1990 (2013).

27. Pradhan, N. R. *et al*. Intrinsic Carrier Mobility of Multi-Layered MoS$_2$ Field-Effect Transistors on SiO$_2$. *Appl. Phys. Lett.* **102**, 123105 (2013).

28. Hwang, W. S. *et al*. Transistors with Chemically Synthesized Layered Semiconductor WS$_2$ Exhibiting $10^5$ Room Temperature Modulation and Ambipolar Behavior. *Appl. Phys. Lett.* **101**, 013107 (2012).

29. Braga, D., Lezama, I. G., Berger, H., & Morpurgo, A. F. Quantitative Determination of the Band Gap of WS$_2$ with Ambipolar Ionic Liquid-Gated Transistors. *Nano Lett.* **12**, 5218-5223 (2012).

30. Fang, F. F., & Fowler, A. B. Transport Properties of Electrons in Inverted Silicon Surfaces. *Phys. Rev.* **169**, 619 (1968).

31. Hartstein, A., Ning, T. H., & Fowler, A. B. Electron Scattering in Silicon Inversion Layers by Oxide and Surface Roughness Original Research. *Surf. Sci.* **58**, 178 (1976).

32. Hasegawa, H., Sawada, T. On the Distribution and Properties of Interface States at Compound Semiconductor-Insulator Interfaces. *Surf. Sci.* **98**, 597 (1980).

33. Ando, T. Screening Effect and Quantum Transport in a Silicon Inversion Layer in Strong Magnetic Fields. *J. Phys. Soc. Jpn.* **43**, 1616-1626 (1977).

34. Mott, N. F. Coulomb Gap and Low-Temperature Conductivity of Disordered Systems. *J. Phys. C: Solid State Phys.* **8**, L239-L240 (1975).

35. Ghatak, S., Pal, A. N., & Ghosh, A. Nature of Electronic States in Atomically Thin $MoS_2$ Field-Effect Transistors. *ACS Nano* **5**, 7707-7712 (2011).

36. Chen, J.-R. *et al*. Control of Schottky Barriers in Single Layer $MoS_2$ Transistors with Ferromagnetic Contacts. *Nano. Lett.* **13**, 3106-3110 (2013).

37. Lang, O., Tomm, Y., Schlaf, R., Pettenkofer, C., & Jaegermann, W. Single Crystalline $GaSe/WSe_2$ Heterointerfaces Grown by Van der Waals Epitaxy. II. Junction Characterization. *J. Appl. Phys.* **75**, 7814 (1994).

38. McDonnel S., *et al*. Hole Contacts on Transition Metal Dichalcogenides: Interface Chemistry and Band Alignments. *ACS Nano* **8**, 6265 (2014).

39. Yang, H. *et al*. Graphene Barristor, a Triode Device with a Gate-Controlled Schottky Barrier. *Science* **336**, 1140 (2012).

40. Ma, N., & Jena, D. Charge Scattering and Mobility in Atomically Thin Semiconductors, *Phys. Rev. X* **4**, 011043 (2014).

41. Yan, J., Zhang, Y., Kim, P., & Pinczuk, A. Electric Field Effect Tuning of Electron-Phonon Coupling in Graphene. *Phys. Rev. Lett.* **98**, 166802 (2007).

42. Yan, J., Henriksen, E. A., Kim, P., & Pinczuk, A. Observation of Anomalous Phonon Softening in Bilayer Graphene. *Phys. Rev. Lett.* **101**, 136804 (2008).

43. Duerloo, K.-A. N., Ong, M. T., & Reed, E. J. Intrinsic Piezoelectricity in Two-Dimensional Materials. *J. Phys. Chem. Lett.* **3**, 2871 (2012).

44. Chuang, H.-J. *et al*. High Mobility $WSe_2$ *p*- and *n*-Type Field-Effect Transistors Contacted by Highly Doped Graphene for Low-Resistance Contacts. *Nano Lett.* **14**, 3594−3601 (2014).

**Acknowledgements**

This work is supported by the U.S. Army Research Office MURI grant W911NF-11-1-0362.

The NHMFL is supported by NSF through NSF-DMR-0084173 and the State of Florida.

**Competing Interests statement**

The authors declare that they have no competing financial interests.

**Authors' Contributions**

LB conceived the project in discussions with NRP, ST, MT and PMA. DR synthesized the $WSe_2$ single crystals. NRP and DR characterized the thickness of the used flakes though AFM techniques. NRP fabricated the field-effect transistors. JMP, DS, MT performed polarized Raman experiments and their dependence on number of layers as well as the corresponding analysis. SF, NPL, ALE and MT have performed Raman measurements as a function of excitation frequency. NRP, SM and LB performed the electrical transport characterization. NRP and LB analyzed the corresponding data. LB wrote the manuscript with the input of all co-authors.

**Correspondence**

Correspondence and requests for materials should be addressed to L.B. (balicas@magnet.fsu.edu)

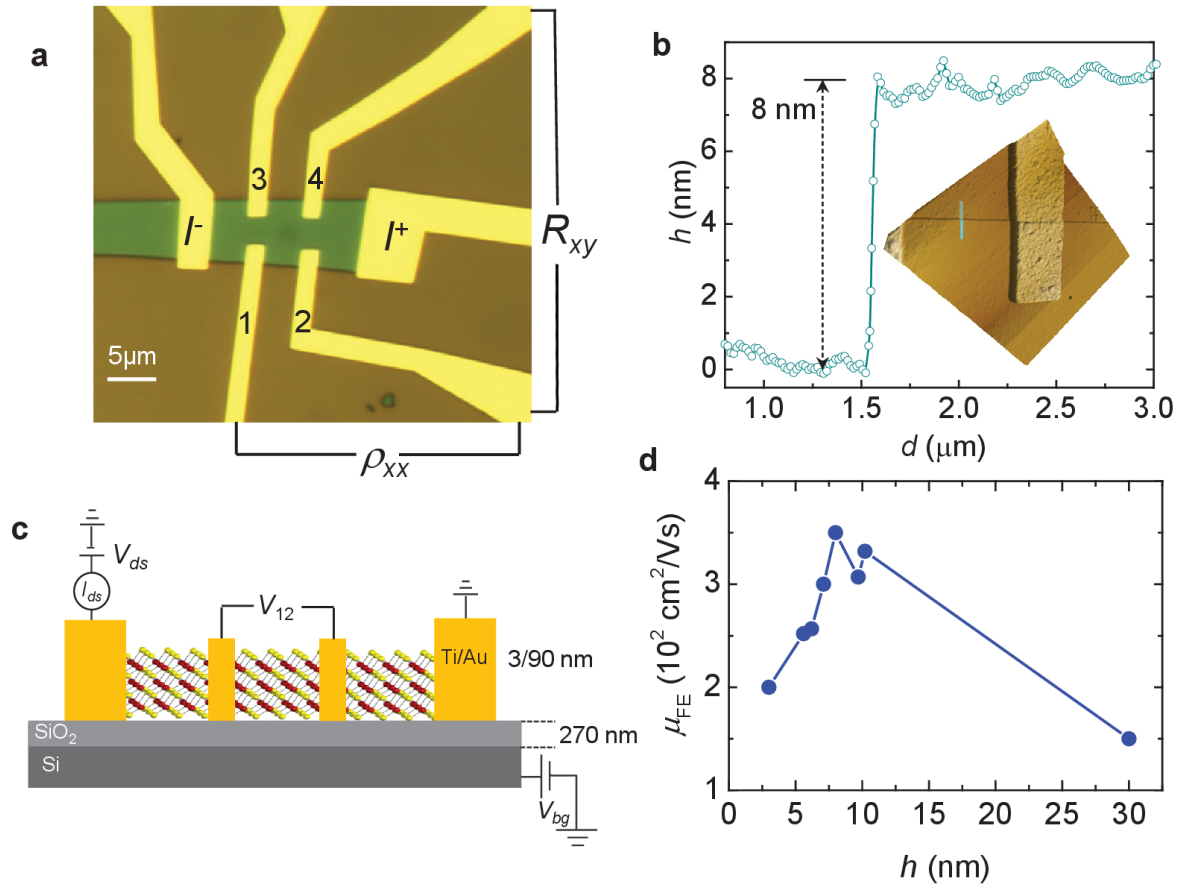

**Figure 1| a** Micrograph of the one of our WSe$_2$ field-effect transistors on a 270 nm thick SiO$_2$ layer on p-doped Si. Contacts, (Ti/Au) used to inject the electrical current ($I_{ds}$), are indicated through labels $I^+$ (source) and $I^-$ (drain), while the resistivity of the device $\rho_{xx}$ was measured through either the pair of voltage contacts labeled as 1 and 2 or pair 3 and 4. The Hall resistance $R_{xy}$ was measured with an AC excitation either through the pair of contacts 1 and 3 or 2 and 4. Length $\ell$ of the channel, or the separation between the current contacts, is $\ell = 15.8$ μm while the width of the channel is $w = 7.7$ μm. **b** Height profile (along the blue line shown in the inset) indicating a thickness of 80 Å, or approximately 12 atomic layers for the crystal in **a**. Inset: atomic force microscopy image collected from a lateral edge of the WSe$_2$ crystal in **a**. **c** Side view sketch of our field-effect transistor(s), indicating that the Ti/Au pads contact all atomic

layers, and of the experimental configuration of measurements. **d** Room temperature field-effect mobility $\mu_{FE}$ as a function of crystal thickness extracted from several FETs based on WSe$_2$ exfoliated onto SiO$_2$. The maximum mobility is observed for ~12 atomic layers

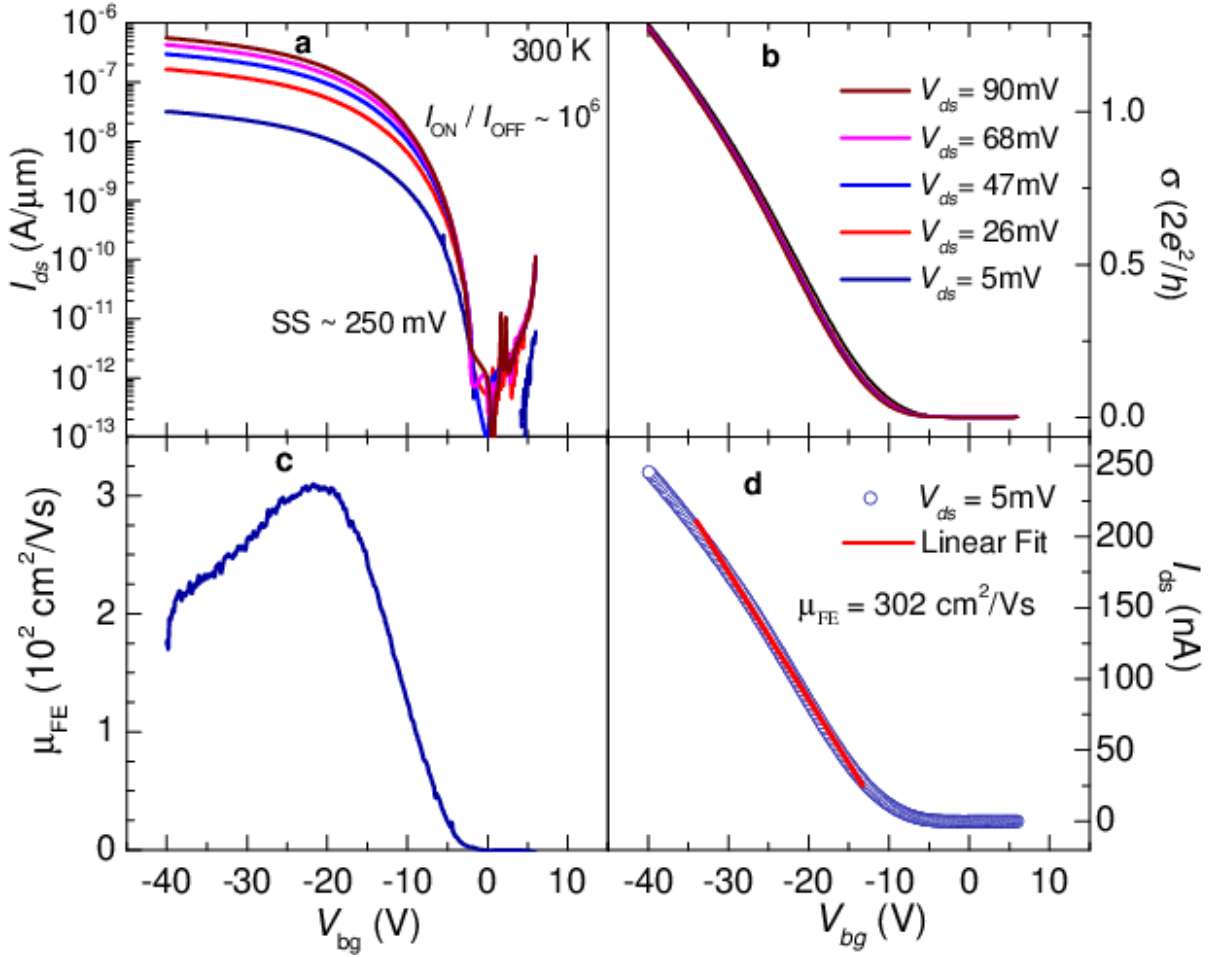

**Figure 2|** **a** Current $I_{ds}$ in a logarithmic scale as extracted from a WSe$_2$ FET at $T = 300$ K and as a function of the gate voltage $V_{bg}$ for several values of the voltage $V_{ds}$, i.e. respectively 5 (dark blue line), 26 (red), 47 (blue), 68 (magenta), and 90 mV (brown), between drain and source contacts. Notice that the ON/OFF ratio approaches $10^6$ and subthreshold swing SS ~250 mV per decade. We evaluated the resistance $R_c$ of the contacts by performing also 4 terminal measurements (see Fig. 7 **a** below) through $R_c = V_{ds}/I_{ds} - \rho_{xx}\ell/w$, where $\rho_{xx}$ is the sheet resistivity

of the channel measured in a four-terminal configuration. We found the ratio $R_c/\rho_{xx} \approx 20$ to remain nearly constant as a function of $V_{bg}$. **b** Conductivity $\sigma = S\ell/w$, where the conductance $S = I_{ds}/V_{ds}$ (from **a**), as a function of $V_{bg}$ and for several values of $V_{ds}$. Notice, how all the curves collapse on a single curve, indicating linear dependence on $V_{ds}$. As argued below, this linear dependence most likely results from thermionic emission across the Schottky-barrier at the level of the contacts. **c** Field effect mobility $\mu_{FE} = (1/c_g\, d\sigma/\, dV_{bg}$ as a function of $V_{bg}$, where $c_g = \varepsilon_r\varepsilon_0/d = 12.789 \times 10^{-9}$ F/cm$^2$ (for a $d = 270$ nm thick SiO$_2$ layer). **d** $I_{ds}$ as a function of $V_{bg}$, when using an excitation voltage $V_{ds} = 5$ mV. Red line is a linear fit whose slope yields a field-effect mobility $\mu_{FE} \approx 300$ cm$^2$/Vs.

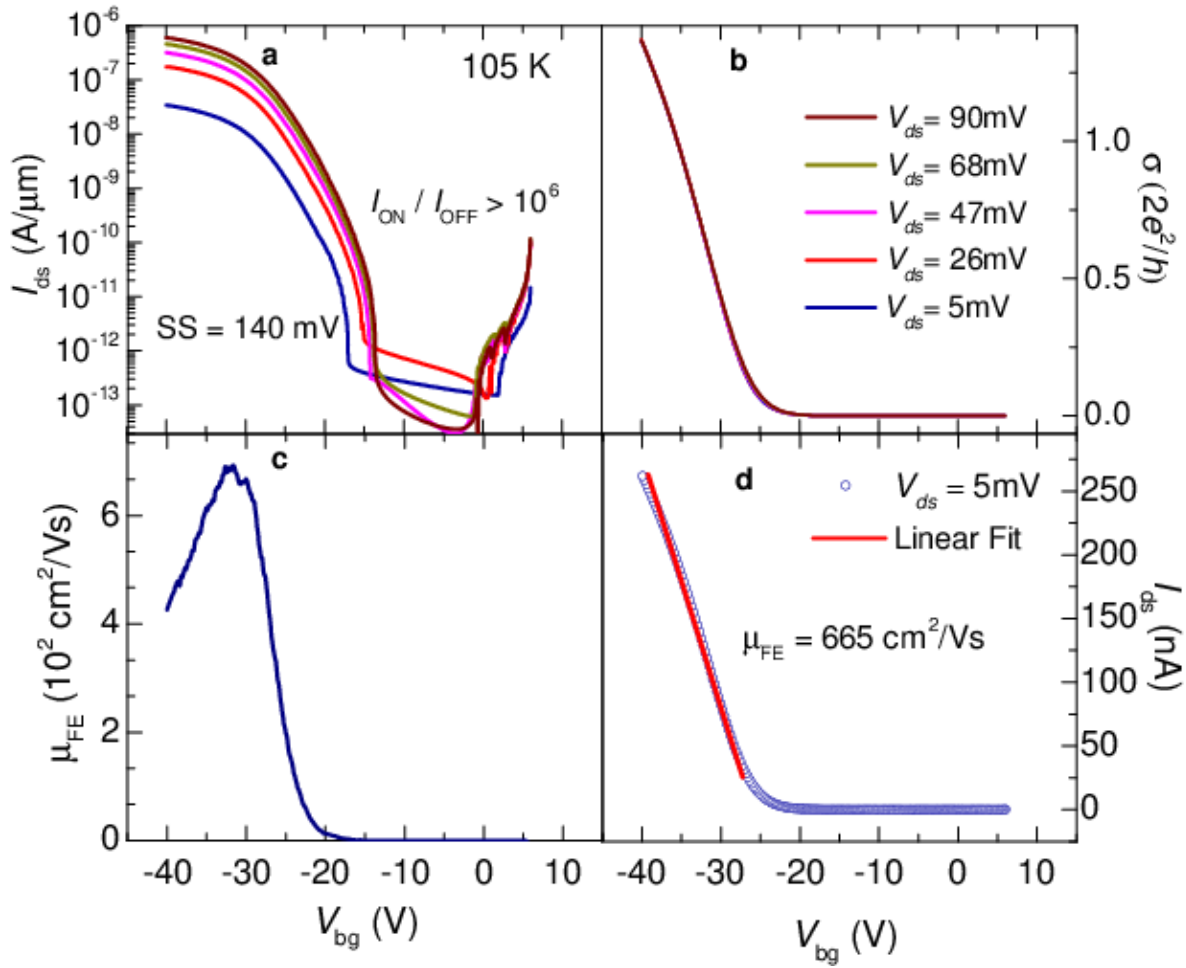

**Figure 3| a** Current $I_{ds}$ in a logarithmic scale as extracted from the same WSe$_2$ FET in Fig. 2 at $T$ = 105 K and as a function of the gate voltage $V_{bg}$ for several values of the voltage $V_{ds}$, i.e. respectively 5 (dark blue line), 26 (red), 47 (magenta), 68 (dark yellow), and 90 mV (brown). Notice that the ON/OFF ratio still approaches $10^6$. **b** Conductivity σ as a function of $V_{bg}$ for several values of $V_{ds}$. Notice that even at lower $T$s all the curves collapse on a single curve. Notice how the threshold gate voltage $V^t_{bg}$ for conduction increases from ~ 0 V at 300 K to ~15 V at 105 K. Below, we argue that the observation of, and the increase of $V^t_{bg}$ as $T$ is lowered, corresponds to evidence for charge localization within the channel. **c** Field effect mobility $\mu_{FE}$= $(1/c_g)\, d\sigma/dV_{bg}$ as a function of $V_{bg}$. **d** $I_{ds}$ as a function of $V_{bg}$, when using an excitation voltage $V_{ds}$ = 5 mV. Red line is a linear fit whose slope yields a field-effect mobility $\mu_{FE} \approx 665$ cm$^2$/Vs.

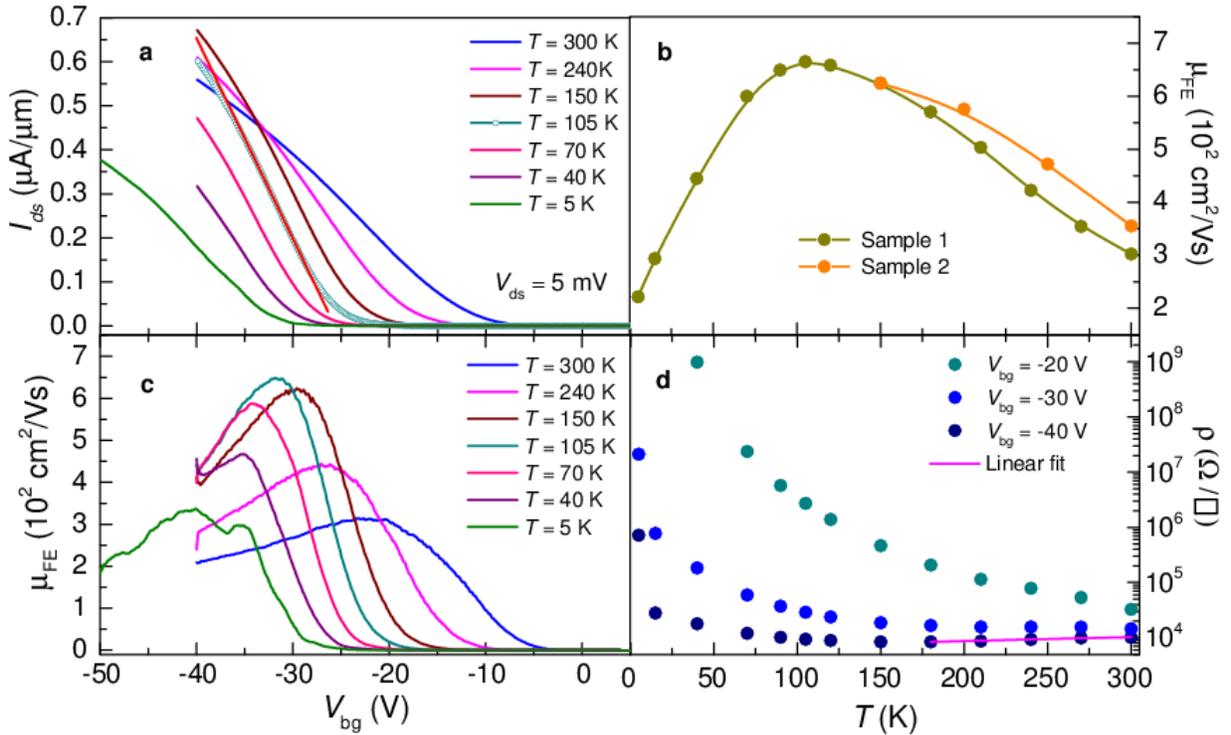

**Figure 4| a** $I_{ds}$ as a function of the gate voltage $V_{bg}$ for several temperatures $T$ and for an excitation voltage $V_{ds}$ = 5 mV. From the slopes of the linear fit (red line) one extracts the

respective values of the field-effect mobility μ$_{FE}$ as a function of the temperature, shown in **b**. Orange markers depicts μ$_{FE}$ for a second, annealed sample. The field-effect mobility is seen to increase continuously as the temperature is lowered down to $T$ = 105 K, beyond which it decreases sharply. **c** μ$_{FE}$ = (1/$c_g$) $d\sigma/dV_{bg}$ as extracted from the curves in **a**. Notice that μ$_{FE}$ still saturates at a value of ≈ 300 cm$^2$/Vs at $T$ = 5 K. **d** Resistivity ρ = 1/σ as a function of $T$ for 3 values of the gate voltage, i.e. -20, -30 and -40 V, respectively (as extracted from the data in **a** or **c**). Magenta line corresponds to a linear fit, describing the behavior of the metallic resistivity, defined by $\partial\rho/\partial T > 0$, observed at higher temperatures when $V_{bg}$ = - 40 V.

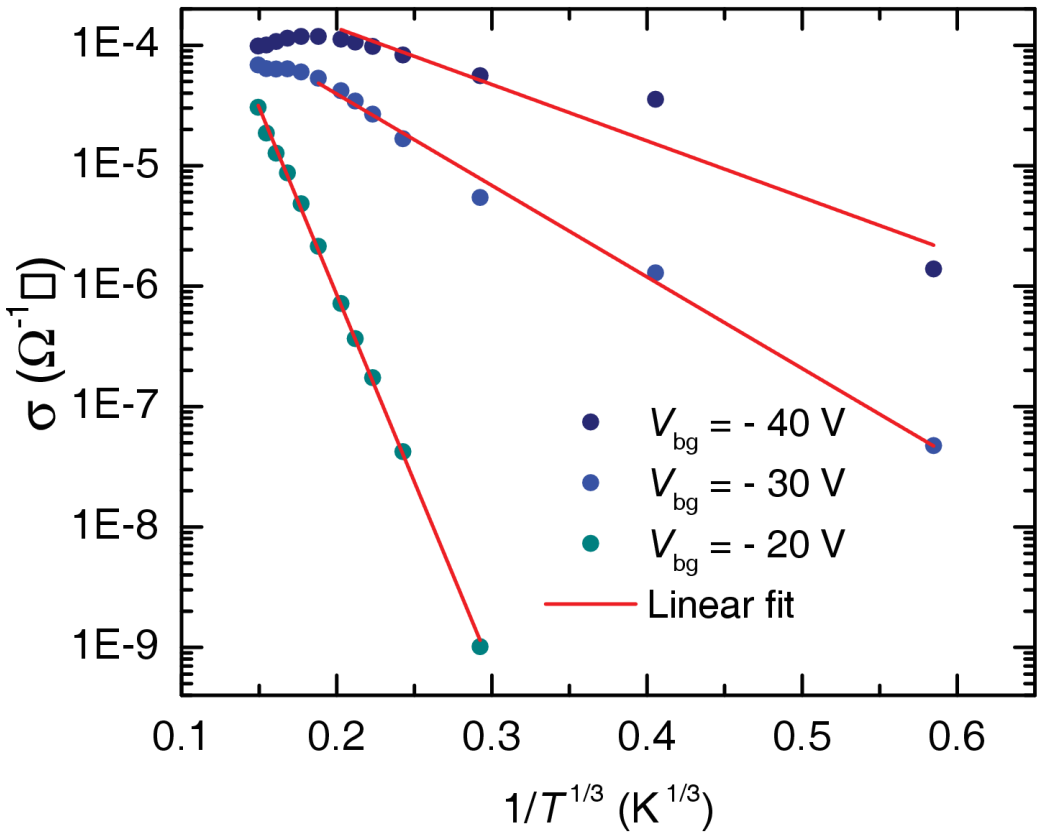

**Figure 5|** Conductivity, i.e. σ = 1/ρ (from the data in Fig. 4 **d**, acquired under $V_{ds}$ = 5 mV) in a logarithmic scale as a function of $T^{-1/3}$. Red lines are linear fits, indicating that at lower $T$s and

for gate voltages below a temperature dependent threshold value $V^t_{bg}(T)$, $\sigma(T)$ follows the dependence expected for two-dimensional variable-range hopping.

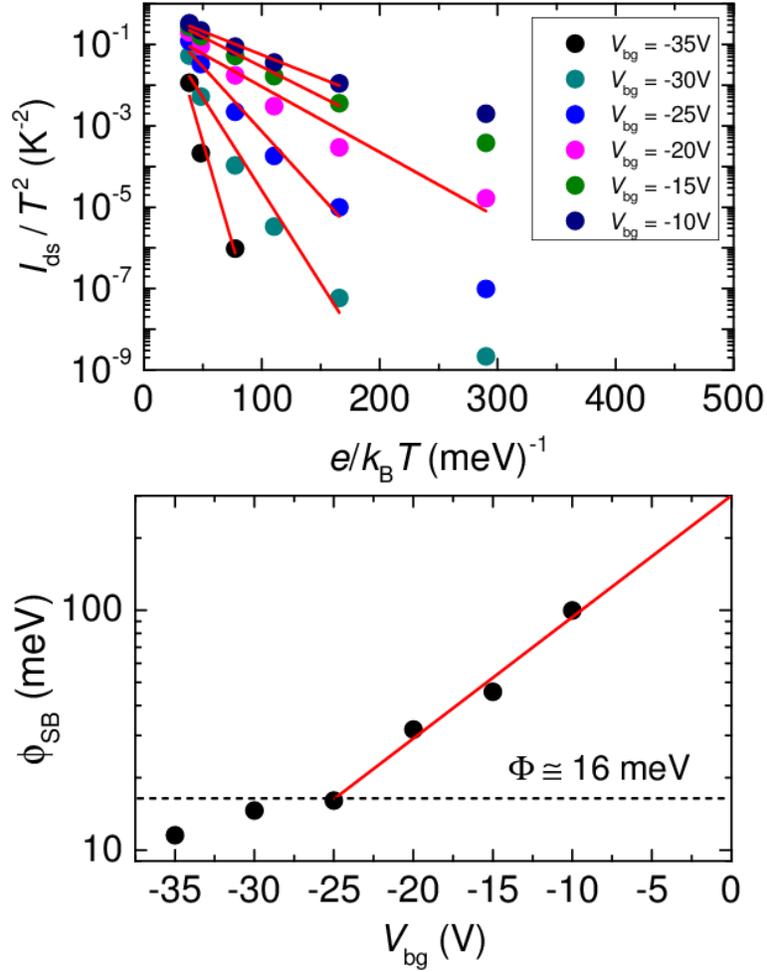

**Figure 6|** Top panel: Drain to source current $I_{ds}$ as a function of $(k_B T/e)^{-1}$ for several values of the gate voltage $V_{bg}$ (from the data in Fig. 4 **a**). Red lines are linear fits from which we extract the value of the Schottky energy barrier $\phi_{SB}$. Bottom panel: $\phi_{SB}$ in a logarithmic scale as a function of $V_{bg}$. Red line is a linear fit. The deviation from linearity indicates when the gate voltage matches the flat band condition[19] from which we extract the size of the Schottky barrier $\Phi \approx 16$ meV.

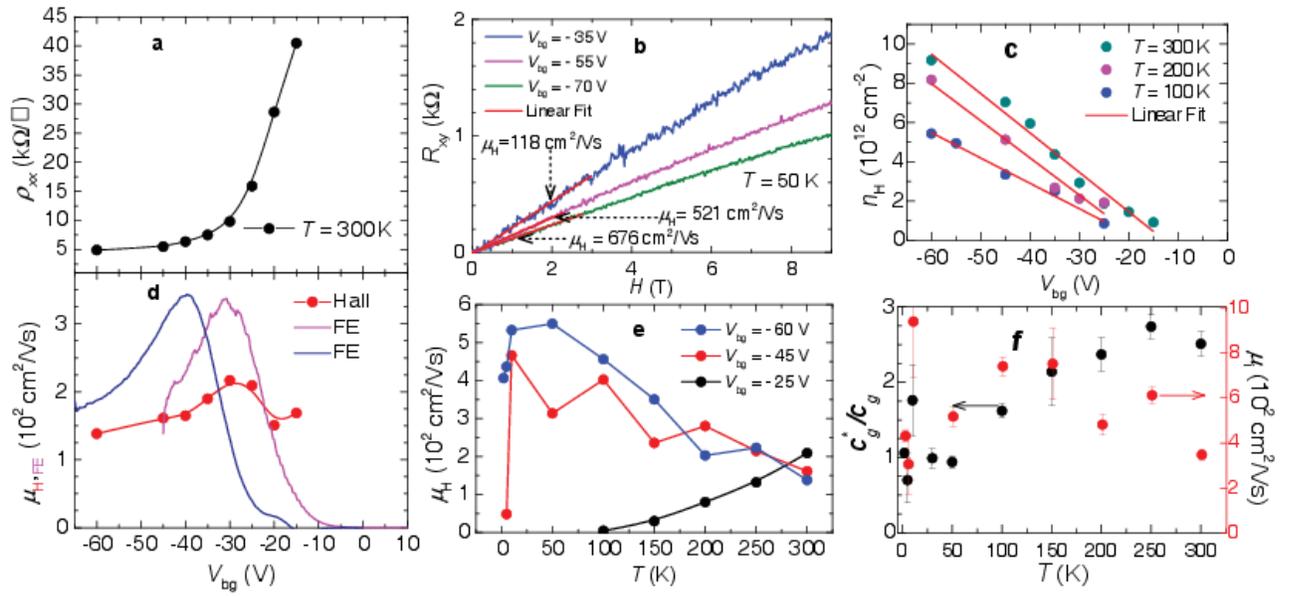

**Figure 7| a** Four-terminal sheet resistance $R_{xx}$ measured at a temperature of $T = 300$ K and as a function of $V_{bg}$ for a second multilayered WSe$_2$ FET after annealing it under vacuum for 24 h. **b** Hall response $R_{xy} = V_H(H)/I_{ds}$ as a function of the external magnetic field $H$, and for several values of the gate voltage $V_{bg}$. Red lines are linear fits from whose slope we extract the values of the Hall constant $R_H(=V_H/HI_{ds})$. **c** Density of carriers $n_H = 1/(eR_H)$ induced by the back gate voltage as a function of $V_{bg}$. Red lines are linear fits from which, by comparing the resulting slope $\sigma = n/V_{bg} = c_g^*/e$ ($c_g^*$ is the effective gate capacitance). **d** Field-effect $\mu_{FE}$ (magenta and blue lines) and Hall $\mu_H = R_H/\rho_{xx}$ (red markers) mobilities (where $\rho_{xx} = R_{xx}w/\ell$, $w$ and $\ell$ are the width and the length of the channel, respectively) as functions of $V_{bg}$ at $T = 300$ K. **e** Extracted Hall mobility $\mu_H$ as a function of $T$ and for several values of $V_{bg}$. $\mu_H$ increases as $T$ is lowered, but subsequently it is seen to decrease below a $V_{bg}$ -dependent $T$. **f** Ratio between experimentally extracted and the ideal, or geometrical gate capacitances $c_g^*/c_g$ (black markers) and the mobilities $\mu_i = c_g^*/c_g\ \mu_H\ (V_{bg}= - 60$ V) (red markers) as functions of $T$. $\mu_i$ are the mobility values that one

would obtain if the gate capacitance displayed its ideal $c_g$ value in absence of spurious charges in the channel.

Supplemental information to manuscript titled: **"Hall and field-effect mobilities in few layered *p*-WSe₂ field-effect transistors"** by Nihar R. Pradhan[1], Daniel Rhodes[1], Shariar Memaran[1], Jean M. Poumirol[1], Dmitry Smirnov[1], Saikat Talapatra[2], Simin Feng, Nestor Perea-Lopez,[3] Ana L. Elias,[4] Mauricio Terrones,[4] Pulickel M. Ajayan,[5] and Luis Balicas[1]

[1]*National High Magnetic Field Laboratory, Florida State University, Tallahassee-FL 32310, USA*
[2]*Physics Department, Southern Illinois University, Carbondale-IL 62901-4401, USA*
[3]*Department of Physics, Department of Materials Science and Engineering and Materials Research Institute, The Pennsylvania State University, University Park, PA 16802, USA*
[4]*Department of Physics, Department of Materials Science and Engineering and Materials Research Institute. The Pennsylvania State University, University Park, PA 16802, USA*
[5]*Department of Mechanical Engineering and Materials Science, Rice University, Houston, TX 77005 USA*

**Current-Voltage characteristics and leakage voltage**

In the right panel of Fig. S1 below, we show the current flowing through the drain-source contacts as a function of the excitation voltage $V_{ds}$ for several values of the back gate voltage. As already inferred from Figs. 2 and 3 in the main text the response of our field-effect transistors is quite linear (as if ohmic) for excitation voltages below 100 mV. The left panel shows an example of the leakage current flowing through the back gate as the back gate voltage (in Fig. 2 within the main text) is swept. As seen, when the current through our FETs surpasses 1 μA, e.g. for $V_{bg} > 40$ V, the leakage current does not even reach 1 nA.

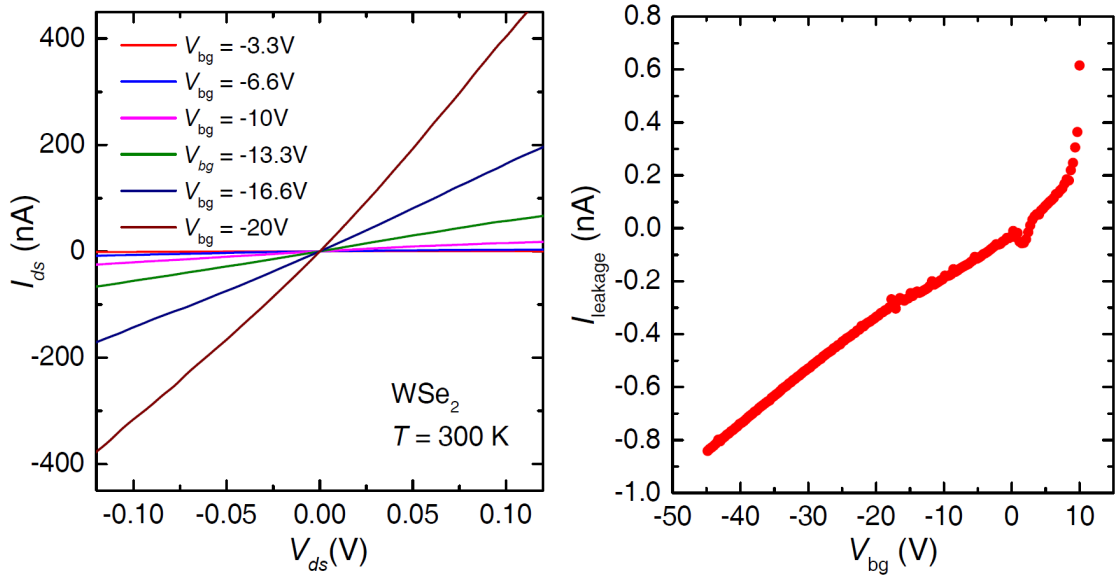

**Figure S1|** Left panel: Drain-source current $I_{ds}$ as a function of the drain-source excitation voltage $V_{ds}$ for several values of the back-gate voltage $V_{bg}$, at room temperature. Right panel: typical leakage current flowing through the back gate as function of the back gate voltage $V_{bg}$ for one of our WSe₂ based FETs.

**Evaluating the Schottky barriers through thermionic emission theory**

When evaluating the Schottky barrier through thermionic emission theory, or through the expression (1) in the main text, i.e. $I_{ds} = AA^*T^2 \exp(e\phi_{SB}/k_BT)$, we have considered the possibility that the low dimensionality of this system might lead to a power in temperature distinct to a $T^2$ dependence, such as a commonly assumed $T^{3/2}$ term, or even the possibility of a hitherto not reported $T$-linear pre-factor. As seen in Fig. S2 below, all three exponents on temperature lead to similar linear fits at higher temperatures. Hence, we cannot unambiguously define the correct power law. One could have left the exponent as a free fitting parameter. However, it would have been difficult to justify theoretically an arbitrary power law.

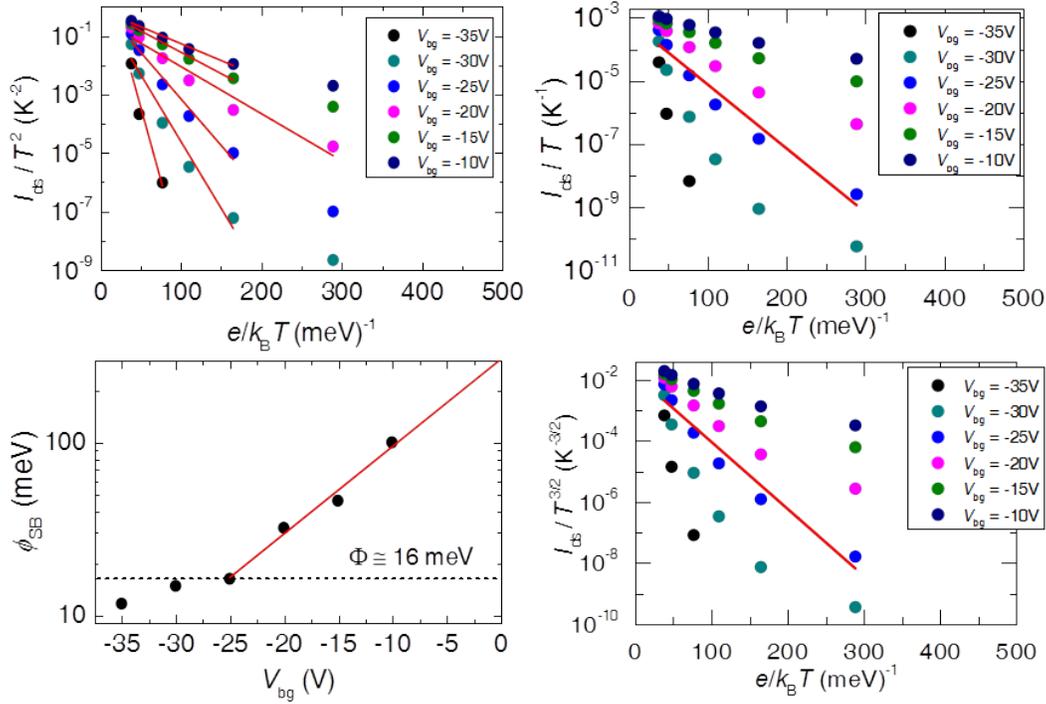

**Figure S2|** Top left panel: Drain to source current $I_{ds}$ as a function of $(k_BT/e)^{-1}$ for several values of the gate voltage $V_{bg}$ (from the data in Fig. 4 **a** in the main text). Red lines are linear fits from which we extract the value of the Schottky energy barrier $\phi_{SB}$. Bottom left panel: $\phi_{SB}$ in a logarithmic scale as a function of $V_{bg}$. Red line is a linear fit. The deviation from linearity would indicate when the gate voltage matches the flat band condition[19] from which one would extract the size of the Schottky barrier $\Phi \approx 16$ meV. Top Right panel: $I_{ds}$ *normalized by temperature* as a function of $(k_BT/e)^{-1}$. Bottom Right panel: $I_{ds}$ normalized by $T^{3/2}$ as a function of $(k_BT/e)^{-1}$. In both panels the red lines are linear fits. Notice how the distinct power laws in temperature lead to linear fits of similar quality.

**Raman Scattering**

In order to understand the extremely high values of mobility shown here, we explored through Raman spectroscopy the quality of the flakes extracted from our single crystals. Here, our goal is to evaluate the width of the Raman peaks since it reflects the coherence and the mean free path of the phonons and therefore the strength of the electron-phonon scattering. A complete Raman study as a function of the number of layers, and laser frequency will be presented elsewhere.

The Raman spectra were measured in a backscattering geometry using a 532.1 nm laser excitation. The laser light was injected into an optical fiber, guiding the excitation to the sample stage. The excitation spot size was about 10 mm in diameter. The scattered light collected by a x100 microscope objective and directed into a collection fiber, and then guided to a spectrometer equipped with a liquid-nitrogen-cooled CCD camera. The spectra were acquired in the spectral region from 150 to 330 cm$^{-1}$ with a spectral resolution of approximately 1 cm$^{-1}$. The peak widths were obtained after correcting for instrumental broadening following the procedure in Ref. [S1]. The Raman spectra shown in Fig. S3 was acquired with an incident laser power of 1.5 mW; we observed that the Raman peaks (position and broadening) are insensitive to the power level when measured with laser power below power densities of 1500 W/cm$^2$.

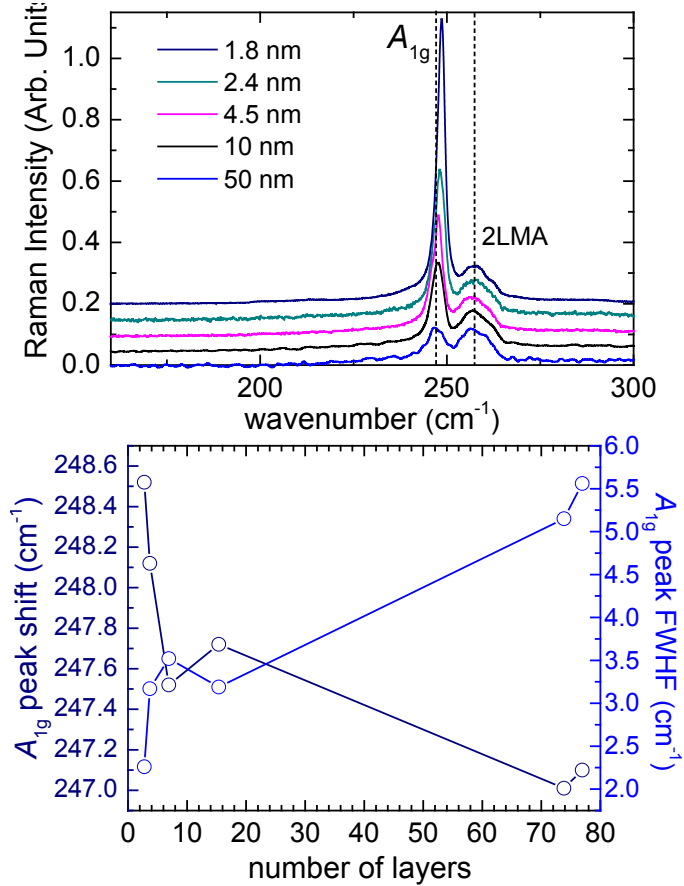

**Figure S3**| Top panel: Intensity of the Raman lines as a function of the wavenumber for several exfoliated WSe$_2$ flakes of distinct thicknesses. The thickness of each flake was determined through atomic force microscopy measurements. Notice how the peak associated with a mixed $E_g+A_{1g}$ mode grows and sharpens as the number of layers is reduced. Lower panel: Position of the aforementioned peak as determined from Lorentzian fits as a function of the number of layers (dark blue markers). Blue markers depict the full width at half maximum (FWHM) of the $E_g+A_{1g}$ peak as a function of the number of layers as resulting from the Lorentzian fits. Notice the shift to higher frequencies and the remarkable increase in sharpness as the number of layers decreases.

Here, we concentrate on the two main peaks observed in the range between 200 cm$^{-1}$ and 300$^{-1}$ when using a laser excitation of 532.1 nm. In a number of previous reports the leftmost peak was identified with the $E^1_{2g}$ shear mode (vibration of the W-Se bond), and the broad feature at its right with the $A_{1g}$ mode (out-of-plane optical vibration of the Se atoms). The $A_{1g}$ mode is known to be sensitive to the polarization of the reflected light relative to the polarization of the incident beam while the $E^1_{2g}$ is not. Through a polarized Raman scattering study we found the leftmost peak to be very sensitive to the relative polarization between the incident and the reflected light. However, it does not become completely inactive under cross-polarization indicating that it corresponds to a mode of mixed ($E + A_{1g}$) character. In effect, in monolayered WSe$_2$, the experimental Raman spectrum exhibits the presence of the perpendicular mode $A'_1$ and the in plane $E'$ almost degenerate at around 250 cm$^{-1}$ [S2, S3]; according to our calculations $A'_1$ is at 250.23 cm$^{-1}$ and the $E'$ at 249.36 cm$^{-1}$ (for details, please see Ref. [S4]). By adding layers, we observed that the out of plane modes $A'_1$ ($A_{1g}$) shift to higher frequencies and the $E'$ ($E_g$) displace to lower frequencies, a behavior that has been reported experimentally by different authors in this and other STMDs (see, Fig. S3 above and also Refs. [S1-S5] ). Experimentally, these modes in the bulk 3-D crystals are associated with the $A_{1g}$ (located at 251 cm$^{-1}$) and the $E_{2g}$ (around 247 cm$^{-1}$) irreducible representations of the $D_{6h}$ point group, exhibiting inversion symmetry. In fact, for a larger number of layers we collected a rather complex Raman spectrum for WSe$_2$ with several additional higher-order mixed modes. Currently, we are performing additional calculations and measurements to understand this complex phonon spectrum which is similar to data collected by other groups [6], but at first glance looks *distinct* from the MoS$_2$ one [7].

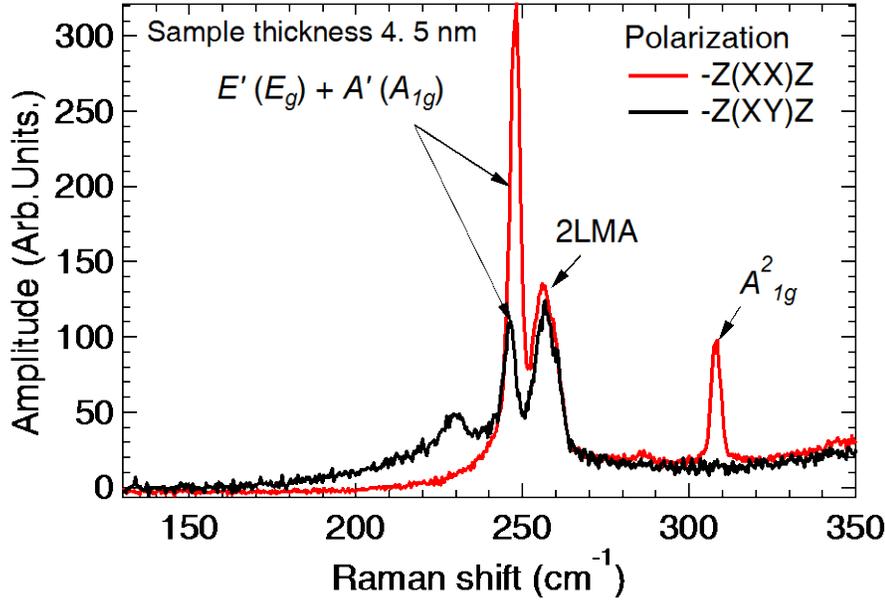

**Figure S4|** Cross polarized Raman spectra for 4.5 nm thick WSe$_2$ flake. Notice how a sharp peak becomes undetectable under cross polarization indicating that it corresponds to out-of-plane lattice vibration modes. According to the calculations in Ref. [S4] this peak corresponds to the $A^2_{1g}$ mode. Finally, notice how the most pronounced peaks are suppressed, although not entirely by the cross polarization, suggesting that they correspond to mixed in-plane ($E'$ or $E_g$) and out of plane modes ($A'$ or $A_{1g}$). We indexed them based on the calculations of Ref. [S4].

In Fig. S3 we show the Raman spectra as a function of the wavenumber as extracted from several flakes of varying thicknesses. The thickness of the number of layers was determined through AFM measurements. As seen in Fig. 2 the mixed $E'$ ($E_g$) + $A'$ ($A_{1g}$) mode red shifts to higher wave-numbers as the number of layers decrease, as predicted and observed for the $E^1_{2g}$ mode in MoS$_2$ [S7, S8, S9]. Remarkably, this shift starts in multi-layered flakes and becomes quite pronounced as the number of layers decrease. Most importantly and as seen in Fig. S3, the height of this $E'$ ($E_g$) + $A'$ ($A_{1g}$) peak increases considerably as the number of layers decrease. It also sharpens, i.e. by a factor greater than 2 with its full width at half maximum (FWHM) decreasing to a very small value of $\cong$ 2.2 cm$^{-1}$ indicating a high in-plane crystallinity for the flakes which does not deteriorate as the number of layers decreases. On the contrary, it suggests an increase in phonon coherence as the number of layers is reduced due to a higher level of crystallinity, possibly due to the removal of defects such as stacking faults or stacking disorder. Most importantly, a considerably large phonon coherence-length necessarily implies small electron-phonon scattering or very weak electron-phonon coupling. This might be an ingredient contributing to the larger mobilities observed here for the WSe$_2$ field-effect transistors. The height and position of the broad 2LMA mode is independent on the number of layers further indicating that the crystallinity of the flakes does not decrease with decreasing the number of layers.

In Fig. S4 above, we demonstrate that the most prominent Raman modes in WSe$_2$ are mixed modes, i.e. composed of both in-plane and out-of-plane vibrational modes. We speculate that the nature of the electron/hole-phonon scattering in WSe$_2$ might be somewhat distinct from the other transition metal dichalcogenides, explaining perhaps its higher carrier mobility at room temperature. This might also lead to a more pronounced effect of the gate voltage on the electron phonon coupling.

**References**

[S1] Tanabe, K., & Hiraishi, J. Experimental-determination of true Raman linewidths from measurements of linewidths observed at different slit openings. *Appl. Spectrosc.* **1981** 35, 436.

[S2] Zhao, W., Ghorannevis, Z., Amara, K. K., Pang, J. R., Toh, M., Zhang, X., Kloc, C., Tan, P. H. and Eda, G. Lattice dynamics in mono- and few-layer sheets of WS$_2$ and WSe$_2$. *Nanoscale* **2013,** 5, 9677-9683.

[S3] Tonndorf, P, Schmidt, R., Bottger, P., Zhang, X., Borner, J., Liebig, A., Albrecht, M., Kloc, C., Gordan, O., Zahn, D. R. T., de Vasconcellos, S. M., Bratschitsch, R. Photoluminescence emission and Raman response of monolayer MoS$_2$, MoSe$_2$, and WSe$_2$. *Opt. Express* **2013**, 1, 4908-4916.

[S4] Terrones, H., Del Corro-Garcia, E., Feng, S., Poumirol, J. M., Smirnov, D., Rhodes, D. Pradhan, N.R., Zhong, L., Nguyen, M.A.T., Elías, A. L., Mallouk, T. E., Balicas, L., Pimenta, M., Terrones, M. New First Order Raman Active Modes in Few Layered Transition Metal Dichalcogenides. *Sci. Rep.* **2014**, 4, 4215.

[S5] Zhang, X., Han, W. P., Wu, J. B., Milana, S., Lu, Y., Li, Q. Q., Ferrari, A. C. & Tan, P. H. Raman spectroscopy of shear and layer breathing modes in multilayer MoS$_2$, *Phys. Rev. B* **2013**, 87, 115413.

[S6] Li, H, Lu, G., Wang, Y., Yin, Z., Cong, C., He, Q., Wang, L., Ding, F., Yu, T., & Zhang, H., Mechanical Exfoliation and Characterization of Single- and Few-Layer Nanosheets of WSe$_2$, TaS$_2$, and TaSe$_2$. *Small* **2013**, 9, 1974–1981.

[S7] Lee, C., Yan, H., Brus, L. E., Heinz, T. F., Hone J., & Ryu, S. Anomalous Lattice Vibrations of Single- and Few-Layer MoS$_2$. *ACS Nano* **2010**, 4, 2695.

[S8] Zeng, H., Zhu, B., Liu, K., Fan, J., Cui, X., & Zhang, Q. M. Low-frequency Raman modes and electronic excitations in atomically thin MoS$_2$ films, *Phys. Rev. B* **2012**, 6, 241301(R).

[S9] Molina-Sanchez, A. & Wirtz, L. Phonons in single-layer and few-layer MoS$_2$ and WS$_2$. *Phys. Rev B* **2011**, 84, 155413.